\documentclass[journal,10pt]{IEEEtran}
\usepackage{graphicx}
\usepackage{caption,color}
\usepackage{subcaption}
\usepackage{refstyle}
\usepackage{amsmath}
\usepackage{algorithm}
\usepackage{algorithmic}
\usepackage{epstopdf}
\usepackage{cite}

\ifCLASSINFOpdf
\else
\fi
\begin{document}
\linespread{1}
\title{Buffer-Aided Relay Selection Algorithms for Physical-Layer Security in Wireless Networks }

\author{Xiaotao Lu and Rodrigo C. de~Lamare
\thanks{Xiaotao Lu is with the Communications Research Group, Department of Electronics, University of York, YO10 5DD York,
U.K., R. C. de Lamare is with CETUC, PUC-Rio, Brazil and with the
Communications Research Group, Department of Electronics, University
of York, YO10 5DD York, U.K. e-mails: xtl503@york.ac.uk;
rodrigo.delamare@york.ac.uk.}}

\maketitle
\begin{abstract}
In this work, we consider the use of buffer-aided relays, linear
precoding techniques and multiple antennas for physical-layer
security in wireless networks. We develop relay selection algorithms
to improve the secrecy-rate performance of cooperative multi-user
multiple-antenna wireless networks. In particular, we propose a
novel finite buffer-aided relay selection algorithm that employs the
maximum likelihood (ML) criterion to select sets of relays which
fully exploit the flexibility offered by relay nodes equipped with
buffers. Numerical results show the benefits of the proposed
techniques as compared to prior art.\\

\end{abstract}

\section{Introduction}

In wireless networks, security is always an aspect of fundamental
importance. Early work suggested the idea of generating keys to
ensure the transmission security \cite{Soni12}. The keys generated
by these algorithms implemented in the network layer are nearly
unbreakable, but the complexity is an obvious drawback. Recently,
physical-layer security has been advocated as a promising
alternative which can ensure the transmission security with much
lower computational complexity and delay.

Relaying techniques for physical-layer security in wireless networks have
attracted significant research interest in the last few years due to their
ability to improve the level of secrecy in the system
\cite{Dong10},\cite{Wang13}. In a typical link-level setup, a group of relays
are used to help the communications between source and destination \cite{lrsc}.
The most reliable relay can be chosen to transmit the signal
\cite{sm_ce},\cite{sicdma},\cite{armo}. Based on this idea, different relay
selection policies can be employed according to different scenarios and
requirements. According to \cite{Krikidis12}, max-min relay selection is
considered as the optimal selection scheme for conventional decode-and-forward
(DF) relay setups. The work in \cite{Krikidis12} introduced max-link relay
selection schemes \cite{tds,jpais,jrdpa} which relax the limitation that the
source and relay transmission must be fixed, and allow each slot to be
allocated dynamically to the source or a relay transmission. In
\cite{Krikidis12} and \cite{Ikhlef12} a max-max relay selection has been
studied with a single destination, where a cooperative network with buffers of
finite size at the relay nodes is considered. The use of a buffer-aided relay
system for improving the physical-layer security has been considered in
\cite{Chen14}, where a max-ratio relay selection is proposed based on the best
channel ratio of the legitimate links to the eavesdroppers. In \cite{Huang14},
a two-hop buffer-aided relay selection has been introduced for physical-layer
security. In the presence of an arbitrary number of users, linear precoding
techniques \cite{Geraci12} can be employed to assist the buffer-aided relay
systems \cite{Liu13}-\cite{Zlatanov13} and improve their secrecy rate.

In this work, we propose novel relay selection strategies for physical-layer
security in buffer-aided multiuser multiple-antenna relay networks
\cite{mmimo_tut}. The first proposed buffer-aided maximum-likelihood (ML) relay
selection (ML-RS) algorithm employs the ML criterion for relay selection. In
each slot, the relay with the best performance is selected to receive or
forward signals. In the presence of multiple relays, different sets of relay
combinations can be employed to further enhance the performance of the network.
Therefore, we also present a technique to select a set of relays based on the
ML criterion which is denoted ML-SRS algorithm. The proposed relay selection
algorithms are compared with existing techniques via computer simulations.

This paper is organized as follows. Section II details the system
model, describes the use of the amplify-and-forward (AF) relaying
protocol in the chosen system, and states the problem of
physical-layer security. Section III presents the proposed ML-type
relay selection algorithm, whereas Section IV details the proposed
ML-type algorithm for selecting sets of relays. Section V shows and
discusses then numerical results, while the conclusions are drawn in
Section VI.

\section{System Model and Physical-Layer Security}

In this section, the system model encompassing a buffer-aided
multiuser multiple-antenna relay network is described along with the
use of the AF relaying protocol and the problem of physical-layer
security in such a system is formulated.

\subsection {System Model}

\begin{center}
\begin{figure}[h]
\includegraphics[scale=0.7]{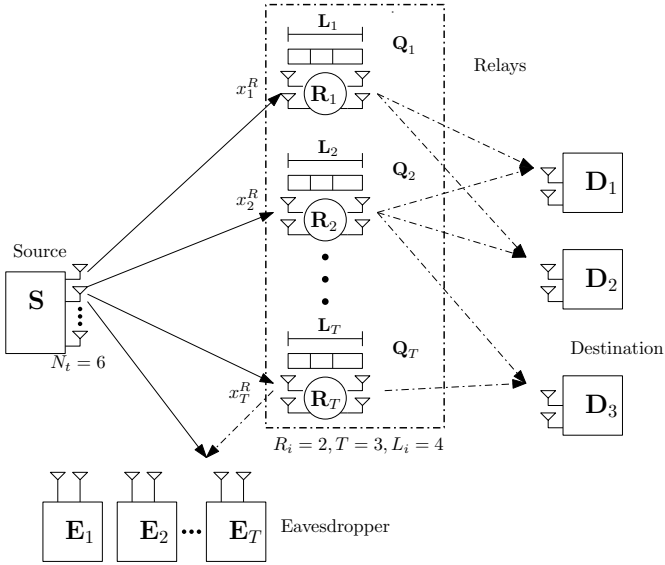}
\caption{Buffer-aided multiuser MIMO system.} \label{fig:Relays}
\end{figure}
\end{center}

We consider a multiuser  multiple-input multiple-output (MIMO)
system with one source $S$ equipped with $N_{t}$ antennas, $N_{D}$
users each with $N_{r}$ antennas at the Destination $D$, a cluster
$\zeta$ with $M$ amplify-and-forward (AF) relays as well as $N_{E}$
eavesdroppers each with $N_{e}$ antennas. All nodes are
characterized by the half-duplex constraint and each relay $R_{m}$
has a buffer $Q_{m}$ with finite size $T$ equipped with $N_{m}$
antennas. The condition $0 \leq \varphi{(Q_{m})}\leq T$ gives the
number of data symbols stored in the buffer $Q_{m}$. With a feedback
channel the selected relay nodes can be informed for further
transmission. Moreover, in order to apply linear precoding
techniques there are two assumptions. One is that the number of
transmit antennas is larger than the sum of the receive antennas.
Another is that the channels are static over a packet of $N_{t}$
symbols. The operation of the system can be divided into two
transmission phases.

In Phase I $(S\rightarrow R)$, the source transmits the signal to
the relays. The transmit signal to the $m$th relay node
${\boldsymbol x}_{sr_{m}} \in {\mathbf{C}}^{N_{m}\times 1}$ with
linear precoding techniques is given by
\begin{equation}
{\boldsymbol x}_{sr_{m}}={\boldsymbol P}_{sr_{m}} {\boldsymbol s}_{sr_{m}},
\label{eqn:xr1}
\end{equation}
where ${\boldsymbol s}_{sr_{m}}\in {\mathbf{C}}^{N_{m}\times 1}$ is
a vector with the transmitted symbols and ${\boldsymbol
P}_{sr_{m}}\in {\mathbf{C}}^{N_{t}\times N_{m}}$ is the precoding
matrix. Considering a linear zero-forcing precoding technique
\cite{Geraci12}, the precoding matrix for the $m$th relay node can
be generated by
\begin{equation}
{\boldsymbol P}_{sr_{m}}={\boldsymbol H}_{sr_{m}}^{H}({\boldsymbol
H}_{sr_{m}} {\boldsymbol H}_{sr_{m}}^{H})^{-1}, \label{eqn:Pr1}
\end{equation}
where ${\boldsymbol H}_{sr_{m}}\in {\mathbf{C}}^{N_{m}\times N_{t}}$
is the channel between the source to the $m$th relay. The received
signal at the $m$th relay node can be expressed as
\begin{equation}
{\boldsymbol y}_{sr_{m}}=\alpha_{sr_{m}}\beta_{sr_{m}}{\boldsymbol H}_{sr_{m}}{\boldsymbol x}_{sr}+{\boldsymbol n}_{sr_{m}},
\label{eqn:ysr}
\end{equation}
where $\alpha_{sr}$ is the power path loss, $\beta_{sr}$ is the
log-normal shadowing (LNS) fading channel loss and ${\boldsymbol
n}_{sr_{m}} \in {\mathbf{C}}^{N_{m}\times 1}$ is the Gaussian noise
between the source and the relay which follows the distribution
$\mathcal{CN}(0,\sigma_{n_{sr}}^{2})$. According to \cite{Thomas12},
$\alpha_{sr}$ is known as the distance based fading (or path loss).
It is a representation of how a signal is attenuated the further it
travels in the medium the system operates within. An exponential
based path loss model can be described by
\begin{equation}
\alpha=\frac{\sqrt{L}}{\sqrt{d^{\rho}}}
\label{eqn:alpha}
\end{equation}
where $L$ is the known path loss associated with $D$, $d$ is the
distance of interest relative to $D$ and $\rho$ is the path loss
exponent, which is typically set between $2$ and $5$. The parameter
$\beta_{sr}$ refers to the shadow fading which can be described
using the log-normal probability distribution given by
\begin{equation}
\beta=10^{(\frac{\sigma_{s}\mathcal{CN}(0,1)}{10})}
\label{eqn:beta}
\end{equation}
where $\sigma_{s}$ is the shadowing spread in dB which is typically
given between $0$ and $9$dB \cite{Thomas14}. Both
$\alpha_{sr}$ and $\beta_{sr}$ are used to describe the position of
the relay.

\subsection{Amplify-and-Forward (AF) Relaying Protocol}

In Phase II $(R\rightarrow D)$, the relays transmit signals to the
destination with the AF relaying protocol. Each relay will transmit
a weighted version of the noisy signal that they received during
Phase I. Let the transmitted signal of all relays be denoted by the
product $\rm{diag}\{\boldsymbol w\}\boldsymbol y_{sr}$. To simplify
the description of the system, we assume that each relay node and
user has the same number of antennas $N_{m}=N_{r}$ and the weight at
each relay node $\rm{diag}\{\boldsymbol w\}=\boldsymbol I$. Then the
received signal at the destination for the $r$th user ${\boldsymbol
y}_{d_{r}} \in {\mathbf{C}}^{N_{r}\times 1}$ can be expressed as
\begin{equation}
{\boldsymbol y}_{d_{r}}={\alpha_{d_{r}}\beta_{d_{r}}{\boldsymbol
H}_{d_{r}} {\boldsymbol P}_{d_{r}}{\boldsymbol
y}_{sr_{m}}}+{\boldsymbol n}_{d_{r}}, \label{eqn:yrd}
\end{equation}
where ${\boldsymbol H}_{d_{r}} \in {\mathbf{C}}^{N_{r}\times (M*N_{m})}$ and
${\boldsymbol P}_{d_{r}} \in {\mathbf{C}}^{(M*N_{m})\times N_{r}}$ are the
channel and the precoding matrices \cite{lcbd,gb,mbthp_conf,mbthp,rmbthp} from
the relays to the destination, where advanced estimation
\cite{mimojio,stjio,jidf,mberjio,locsme} and detection
\cite{itic,spa_df,mfdf,dfcc,mbdf,did} algorithms can be used . The vector
${\boldsymbol n}_{d_{r}} \in {\mathbf{C}}^{N_{r}\times 1}$ represents the
Gaussian noise at the destination.

\subsection{Physical-Layer Security}

According to information theory, the level of secrecy is measured by
the uncertainty of Eve about the message $R_{e}$ which is called the
equivocation rate. The secrecy capacity $C_{s}$ is the supremum of
all achievable secrecy rates.

The MIMO system secrecy capacity can be expressed as:
\begin{equation}
\begin{split}
C_{s} &= \max_{{\boldsymbol Q}_{s}\geq 0, \rm Tr({\boldsymbol
Q}_{s})
 = E_{s}}\log(\det({\boldsymbol I} +{\boldsymbol H}_{ba} {\boldsymbol Q}_{s} {\boldsymbol H}_{ba}^H))\\
& \quad ~~~~~~~~~~~~~~ -\log(\det({\boldsymbol I}+ {\boldsymbol
H}_{ea}{\boldsymbol Q}_{s} {\boldsymbol H}_{ea}^H)),
\end{split}
\label{eqn:Cs1}
\end{equation}
In (\ref{eqn:Cs1}) ${\boldsymbol Q}_{s}$ is the covariance matrix
associated with the signal after precoding, whereas ${\boldsymbol
H}_{ba}$, ${\boldsymbol H}_{ea}$ represent the channel from source
$a$ to user $b$, eavesdropper $e$, respectively. Then based on
(\ref{eqn:Rs1}), the capacity from the source to the relay can be
expressed as:
\begin{equation}
C_{sr} =\log(\det({\boldsymbol I}+{\boldsymbol H}_{sr} {\boldsymbol
Q}_{s} {\boldsymbol H}_{sr}^H)). \label{eqn:Rs1}
\end{equation}
Similarly, the capacity to the eavesdropper is given by
\begin{equation}
C_{se} =\log(\det({\boldsymbol I}+{\boldsymbol H}_{se} {\boldsymbol Q}_{s} {\boldsymbol H}_{se}^H))
\label{eqn:Re1}
\end{equation}
And according to (\ref{eqn:yrd}), the rate at the destination can be
expressed as:
\begin{equation}
R_{d} = \frac{1}{2}\log(\det({\boldsymbol I}+\frac{{\boldsymbol
H}_{rd}\boldsymbol P_{d}{\boldsymbol H}_{sr} {\boldsymbol Q}_{s}
{\boldsymbol H}_{sr}^H\boldsymbol P_{d}^{H}{\boldsymbol
H}_{rd}^H}{{\boldsymbol H}_{rd}{\boldsymbol Q}_{s}{\boldsymbol
H}_{rd}^H+\boldsymbol I})), \label{eqn:Rd1}
\end{equation}
where the scalar factor $1/2$ is due to the fact that two time units
are required in two phases. Similarly, the rate at the eavesdropper
is given by
\begin{equation}
R_{e} =\frac{1}{2}\log(\det({\boldsymbol \Gamma}+\frac{{\boldsymbol H}_{re}\boldsymbol P_{d}{\boldsymbol H}_{sr} {\boldsymbol Q}_{s} {\boldsymbol H}_{sr}^H\boldsymbol P_{d}^{H}{\boldsymbol H}_{re}^H}{{\boldsymbol H}_{re}{\boldsymbol Q}_{s}{\boldsymbol H}_{re}^H+\boldsymbol I}))
\label{eqn:Re1}
\end{equation}
where $\boldsymbol \Gamma=\boldsymbol I + \boldsymbol H_{se}\boldsymbol Q_{s}\boldsymbol H_{se}^{H}$.
Then the overall buffer-aided relay system secrecy rate is given by,
\begin{equation}\label{eqn:Rbf1}
\begin{split}
R&=R_{d}-R_{e}\\
 &=\frac{1}{2}\log(\det({\boldsymbol I}+\frac{{\boldsymbol H}_{rd}\boldsymbol P_{d}{\boldsymbol H}_{sr} {\boldsymbol Q}_{s} {\boldsymbol H}_{sr}^H\boldsymbol P_{d}^{H}{\boldsymbol H}_{rd}^H}{{\boldsymbol H}_{rd}{\boldsymbol Q}_{s}{\boldsymbol H}_{rd}^H+\boldsymbol I}))\\
 &  \quad -\frac{1}{2}\log(\det({\boldsymbol \Gamma}+\frac{{\boldsymbol H}_{re}\boldsymbol P_{d}{\boldsymbol H}_{sr} {\boldsymbol Q}_{s} {\boldsymbol H}_{sr}^H\boldsymbol P_{d}^{H}{\boldsymbol H}_{re}^H}{{\boldsymbol H}_{re}{\boldsymbol Q}_{s}{\boldsymbol H}_{re}^H+\boldsymbol I}))\\
\end{split}
\end{equation}
Then the secrecy rate for each pair of user and relay can be
calculated with a similar formula.

\section{Proposed Buffer-aided ML criterion Based Relay Selection (ML-RS) Algorithm}

According to \cite{Thomas12}, to approach the channel capacity, the
maximum likelihood (ML) decoder can be applied at the receiver to
obtain the following estimate of the transmitted data symbols:
\begin{equation}
{\hat{ \boldsymbol x}}=\rm{arg} \min_{\boldsymbol x \in \phi}\|\boldsymbol y - \boldsymbol H\boldsymbol x\|^{2},
\label{eqn:ml3}
\end{equation}
With a cooperative relay system, to achieve as high transmission
rate as possible, the maximum likelihood (ML) criterion can be used
to select the best relay for transmission which can be expressed as:
\begin{equation}
{\hat{ \boldsymbol H}}=\rm{arg} \min_{\boldsymbol H \in \chi}\|\boldsymbol y - \alpha\beta\boldsymbol H\boldsymbol x\|^{2},
\label{eqn:ml}
\end{equation}
where $\chi$ represents the set of all possible combinations of
links $\boldsymbol H$. For example, if there are $3$ relays in the
relay poll. The total number of possible link combinations would be
$3$. If the total channel $\boldsymbol H$ is formed as $\boldsymbol
H=[\boldsymbol H_{1} \quad \boldsymbol H_{2} \quad\boldsymbol
H_{3}]$. Then the possible selection would be $\boldsymbol H_{1}$,
$\boldsymbol H_{2}$ or $\boldsymbol H_{3}$. In this proposed
algorithm, we use the ML criterion as the relay selection criterion,
then combine the ML relay selection strategy with the buffer. The
selection procedure can be described by
\begin{equation}
{ \boldsymbol R^{k}} =\rm{arg} \min_{\boldsymbol R_{m} \in
\zeta}\{{\bigcup_{\boldsymbol R_{p} \in \zeta:\varphi{(Q_{p})}\neq
T}{\hat{ \boldsymbol H}_{S,\boldsymbol R_{p}}}},
{\bigcup_{\boldsymbol R_{q} \in \zeta:\varphi{(Q_{q})}\neq 0}{\hat{
\boldsymbol H}_{\boldsymbol R_{q},D}}}\} \label{eqn:ml2}
\end{equation}
where ${\bigcup_{\boldsymbol R_{p} \in \zeta:\varphi{(Q_{p})}\neq
T}{\hat{ \boldsymbol H}_{S,\boldsymbol R_{p}}}}$  are the channel
combinations with the smallest Euclidean distance according to
Eq.(\ref{eqn:ml}) at the relay. The parameter ${\bigcup_{\boldsymbol
R_{q} \in \zeta:\varphi{(Q_{q})}\neq 0}{\hat{ \boldsymbol
H}_{\boldsymbol R_{q},D}}}$ refers to the channel combinations
related to the destination. In Eq.(\ref{eqn:ml2}), the first aspect
shown is that if the buffer is full $(\varphi{(Q_{m})}=T)$ the relay
can only transmit signals and if the buffer is empty
$(\varphi{(Q_{m})}=0)$ the relay can only receive signals. Secondly,
as the relay selection policy is combined with the buffer, the
buffer-aided relay can achieve better channel selection over time.
With the ML relay selection (ML-RS) strategy illustrated in
(\ref{eqn:ml2}), the main steps of the proposed Buffer-Aided ML-RS
algorithm are shown in Algorithm 1.

\begin{algorithm}
\caption{Buffer-Aided ML-RS Algorithm}
\begin{algorithmic}
\LOOP
\STATE $\varphi{(Q_{m})}=0$
\FOR {$m=1:M$}
\STATE ${{\hat{ \boldsymbol H_{S,R_{m}}}}}=\rm{arg} \min_{\boldsymbol R_{m} \in \zeta}\|\boldsymbol y_{sr_{m}} - \alpha_{sr_{m}}\beta_{sr_{m}}\boldsymbol H_{sr_{m}}\boldsymbol x_{sr_{m}}\|^{2}$
\ENDFOR
\FOR {$d=M+1:M+N_{D}$}
\STATE ${{\hat{ \boldsymbol H_{\boldsymbol R_{d},D}}}}=\rm{arg} \min_{\boldsymbol R_{d_{r}} \in \zeta}\|\boldsymbol y_{d_{r}} - \alpha_{d_{r}}\beta_{d_{r}}\boldsymbol H_{d_{r}}\boldsymbol y_{sr_{(d-M)}}\|^{2}$
\ENDFOR
\STATE ${ \boldsymbol R^{k}}=\rm{arg} \min_{\boldsymbol R_{k} \in \zeta}$
\STATE $\{{\bigcup_{\boldsymbol R_{m} \in \zeta:\varphi{(Q_{m})}\neq T}{\hat{ \boldsymbol H}_{S,\boldsymbol R_{m}}}}, {\bigcup_{\boldsymbol R_{d} \in \zeta:\varphi{(Q_{d})}\neq 0}{\hat{ \boldsymbol H}_{\boldsymbol R_{d},D}}}\}$
\IF {$k<=M \& \varphi{(Q_{k})}\neq T$}
\STATE $Q_{k}=y_{sr}$
\STATE $\varphi{(Q_{k})}=\varphi{(Q_{k})}+N_{m}$
\ELSE
\IF {$k>M \& \varphi{(Q_{k})}\neq 0$}
\STATE $y_{sr_{(d-M)}}=Q_{k}$
\STATE $\varphi{(Q_{k})}=\varphi{(Q_{k})}-N_{m}$
\ENDIF
\ENDIF
\ENDLOOP
\end{algorithmic}
\end{algorithm}

\section{Proposed Buffer-aided ML criterion with Set of Relays Selection (ML-SRS) Algorithm}

In the presence of an arbitrary number of relays, one can consider
the selection of a set of relays rather than the best relay. The
selection criterion has the same equation as Eq.(\ref{eqn:ml}). We
also take three relays as an example, then the total number of
possible combinations of links would be $7$. If the total channel
$\boldsymbol H$ is formed as $\boldsymbol H=[\boldsymbol H_{1} \quad
\boldsymbol H_{2} \quad\boldsymbol H_{3}]$. Then the possible
selection would be $\boldsymbol H_{1}$, $\boldsymbol H_{2}$,
$\boldsymbol H_{3}$, $[\boldsymbol H_{1} \quad \boldsymbol H_{2}]$,
$[\boldsymbol H_{1} \quad \boldsymbol H_{3}]$, $[\boldsymbol H_{2}
\quad \boldsymbol H_{3}]$ and $[\boldsymbol H_{1} \quad \boldsymbol
H_{2} \quad \boldsymbol H_{3}]$. Based on Eq.(\ref{eqn:ml2}), in
Phase II the selection of the set of relays is achieved with the aid
of a feedback channel from the destination to the relays. A sequence
of bits can be used to indicate whether the relays are switched on
or off. The proposed selection of the set of relays denoted ML-SRS
algorithm can be expressed as
\begin{equation}
{ \boldsymbol R^{K}}=\rm{arg} \min_{\boldsymbol R_{M}  \in
\zeta}\{{\bigcup_{\boldsymbol R_{M} \in \zeta:\varphi{(Q_{M})}\neq
T}{\hat{ \boldsymbol H}_{S,\boldsymbol R_{\hat{M}}}}},
{\bigcup_{\boldsymbol R_{\hat{N}} \in
\zeta:\varphi{(Q_{\hat{N}})}\neq 0}{\hat{ \boldsymbol
H}_{\boldsymbol R_{\hat{N}},D}}}\} \label{eqn:ml3}
\end{equation}
where $K,M,N$ are subsets of $\zeta$. The major difference between
Eq.(\ref{eqn:ml3}) and Eq.(\ref{eqn:ml2}) is that, due to the
increased number of relays, different relays can be selected for
transmission which is known as sets of relays selection (SRS). The
set ${\bigcup_{\boldsymbol R_{M} \in \zeta:\varphi{(Q_{M})}\neq
T}{\hat{ \boldsymbol H}_{S,\boldsymbol R_{\hat{M}}}}}$ indicates
that the channel combinations give the smallest Euclidean distance
with respect to an $M$ set of relays selection from the source to
the relays. The set ${\bigcup_{\boldsymbol R_{\hat{N}} \in
\zeta:\varphi{(Q_{\hat{N}})}\neq 0}{\hat{ \boldsymbol
H}_{\boldsymbol R_{\hat{N}},D}}}$ has the same definition with an
$N$ set of relays links to the destination. Compared to the single
relay selection, SRS gives more flexible selections at the relays,
which contributes to the improvement of the secrecy performance. The
main steps are given in Algorithm 2.
\begin{algorithm}
\caption{Buffer-Aided ML-SRS Algorithm}
\begin{algorithmic}
\FOR {$d=M+1:M+N_{D}$} \STATE ${{\hat{ \boldsymbol H_{
d}}_{\boldsymbol R_{\hat{M}},D}}}$ \STATE $=\rm{arg}
\min_{\boldsymbol H_{\hat{M}d}  \in \chi}\|\boldsymbol y_{\hat{M}d}
- \alpha_{\hat{M}d}\beta_{\hat{M}d}\boldsymbol
H_{\hat{M}d}\boldsymbol x_{\hat{M}d}\|^{2}$ \ENDFOR \STATE ${
\boldsymbol R^{K}}=\rm{arg} \min_{\boldsymbol R_{M} \in \zeta}$
\STATE $\{{\bigcup_{\boldsymbol R_{M} \in \zeta:\varphi{(Q_{M})}\neq
T}{\hat{ \boldsymbol H}_{S,\boldsymbol R_{\hat{M}}}}},
{\bigcup_{\boldsymbol R_{\hat{N}} \in
\zeta:\varphi{(Q_{\hat{N}})}\neq 0}{\hat{ \boldsymbol
H}_{\boldsymbol R_{\hat{N}},D}}}\}$
\end{algorithmic}
\end{algorithm}

\section{Simulation Results}

In the simulation of a single-antenna scenario, we assume that the
weight at each relay node is given by $\rm{diag}\{\boldsymbol
w\}=\boldsymbol I$. The transmitter is equipped with $N_{t}=3$
antennas and each relay node is equipped with one antenna for
receiving or transmitting data. Each user is equipped with a single
antenna and the number of users is set to $N_{D}=3$. At the same
time $N_{E}=3$ eavesdroppers are receiving data from the source. In
the simulation of the multiuser MIMO scenario, the transmitter is
equipped with $N_{t}=6$ antennas and each relay node is equipped
with $N_{m}=2$ antennas for receiving or transmitting. Each user is
equipped with $N_{r}=2$ antennas and the number of users is set to
$N_{D}=3$. At the same time $N_{E}=3$ eavesdroppers are all equipped
with $N_{e}=2$ antennas. In both scenarios, a zero-forcing precoding
technique is implemented at the transmitter and at the relays, and
the secrecy rate is calculated with Eq.(\ref{eqn:Rbf1}).

\begin{center}
\begin{figure}[h]
\includegraphics[scale=0.485]{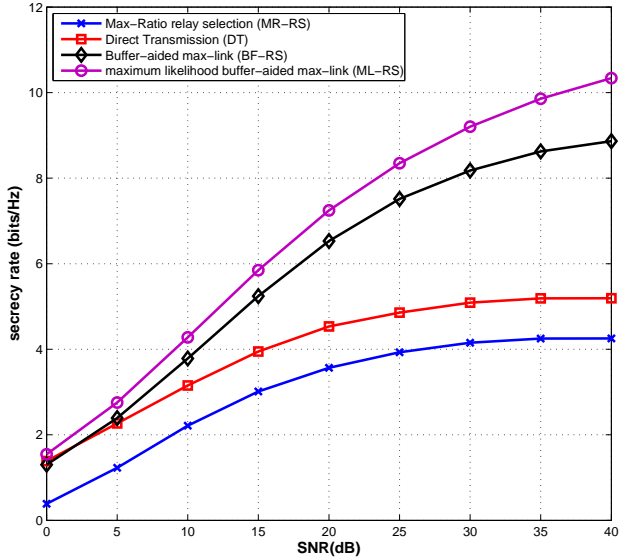}
\caption{Multi-user system with a single-antenna scenario and the
simulation parameters $N_{t}=3, N_{m}=1,N_{r}=1,N_{e}=1$ and
$M=3,N_{D}=3,N_{E}=3,T=3$} \label{fig:buffer1}
\end{figure}
\end{center}

The simulation results shown in Fig. \ref {fig:buffer1} indicate
that the proposed Buffer-Aided ML-RS algorithm outperforms the
conventional relay selection strategy with and without buffers. With
the weight at each relay node set to $\rm{diag}\{\boldsymbol
w\}=\boldsymbol I$, the Max-Ratio relay selection is worse than
Direct Transmission, as the noise is enhanced by the AF relays.
Compared with the Direct Transmission, the buffer-aided max-link
selection policy selects the link with the best performance using
the relays equipped with buffers. In particular, the proposed
Buffer-Aided ML-RS algorithm contributes to the improvement of the
users channel capacity so the secrecy rate will also have an
improvement.

\begin{center}
\begin{figure}[h]
\includegraphics[scale=0.485]{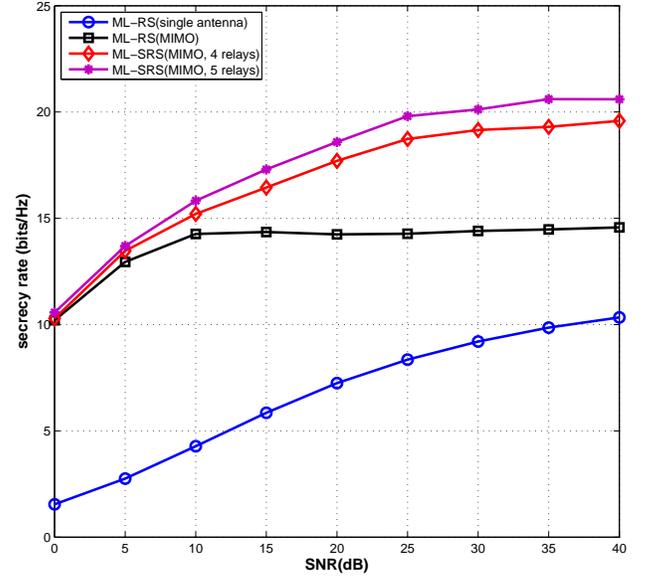}
\caption{Secrecy rate with the ML-SRS strategy for MU-MIMO systems
with $N_{t}=6, N_{m}=2,N_{r}=2,N_{e}=2$ and $M=3,N_{D}=3,N_{E}=3,T=6$}
\label{fig:buffer2}
\end{figure}
\end{center}

In Fig. \ref {fig:buffer2}, we consider a MIMO system scenario in
which the proposed algorithms have better secrecy rate performance
than the single-antenna configuration. In low SNRs, the secrecy rate
of the proposed ML-RS algorithm with the MIMO scenario is around
$8-9bits\/Hz$ better than the single-antenna system. Then the gap
between the two scenario becomes narrower with the increase of the
SNR. Moreover, in the simulation the proposed Buffer-Aided ML-SRS
algorithm with different numbers of relays is compared with the
proposed Buffer-Aided ML-RS algorithm in the MIMO scenario
considered. When we compare the performance of the system with $4$
relays selected with that of the system with $3$ relays, the secrecy
rate is improved significantly. In a scenario with $5$ relays, the
improvement is not as significant as before.

\section{Conclusion}

In this work, we have considered the use of buffer-aided relays,
linear precoding techniques and MIMO for physical-layer security in
wireless networks. We have developed relay selection algorithms to
improve the secrecy-rate performance of cooperative multi-user
multiple-antenna wireless networks. In particular, we have presented
a novel finite buffer-aided relay selection algorithm that employs
the ML criterion to select sets of relays which fully exploit the
flexibility offered by relay nodes equipped with buffers. Numerical
results have shown that the proposed techniques can offer
significant gains in terms of secrecy rate as compared to existing
algorithms.

\bibliographystyle{IEEEtran}
\bibliography{referenceWSA}

\end{document}